\documentclass[draft,grl]{agutex}

\usepackage{graphicx,amssymb,amsmath,bm,ulem}

\authorrunninghead{ANDREJCZUK ET AL.}

\titlerunninghead{OCEAN STOCHASTIC PARAMETRIZATION}

\authoraddr{Corresponding author: L. Zanna,
Atmospheric, Oceanic and Planetary Physics, Department of Physics, University of Oxford, Oxford, OX1 3PU, UK. (laure.zanna@physics.ox.ac.uk)}

\begin{document}

%
%

\title{Oceanic stochastic parametrizations in a seasonal forecast system}

%
%




\authors{M. Andrejczuk\altaffilmark{1}, F. C. Cooper,\altaffilmark{1}, S. Juricke\altaffilmark{1}, T. N. Palmer\altaffilmark{1}, A. Weisheimer \altaffilmark{2}\;\altaffilmark{3} and L. Zanna\altaffilmark{1} }

\altaffiltext{1}{Atmospheric, Oceanic and Planetary Physics, Department of Physics, University of Oxford, Oxford, OX1 3PU, UK.}

\altaffiltext{2}{European Centre for Medium-Range Weather Forecasts, Shinfield Park, Reading, Berkshire, RG2 9AX, UK.}

\altaffiltext{3}{National Centre for Atmospheric Science, Department of Physics,
Atmospheric, Oceanic and Planetary Physics, Oxford University, Oxford, UK.}


%
%


\begin{abstract}

 We study the impact of three stochastic parametrizations in the ocean component of a coupled model, on forecast reliability over seasonal timescales.  The relative impacts of these schemes upon the ocean mean state and ensemble spread are analyzed. The oceanic variability induced by the atmospheric forcing of the coupled system is, in most regions, the major source of ensemble spread.  The largest impact on spread and bias came from the
 Stochastically Perturbed Parametrization Tendency (SPPT) scheme  -
 which has proven particularly effective in the atmosphere. 
 The key regions affected are eddy-active regions, namely the
 western boundary currents and the
 Southern Ocean.  However, unlike its impact in the atmosphere, SPPT in the ocean
 did not result in a significant decrease in forecast 
 error.  Whilst
 there are good grounds for implementing stochastic schemes in ocean
 models, our results suggest that they will have to be more
 sophisticated. Some
 suggestions for next-generation stochastic schemes are made.

 
\end{abstract}

%
%

%

\begin{article}

%
%

\section{Introduction}

\begin{figure*}
\noindent\includegraphics[width=40pc]{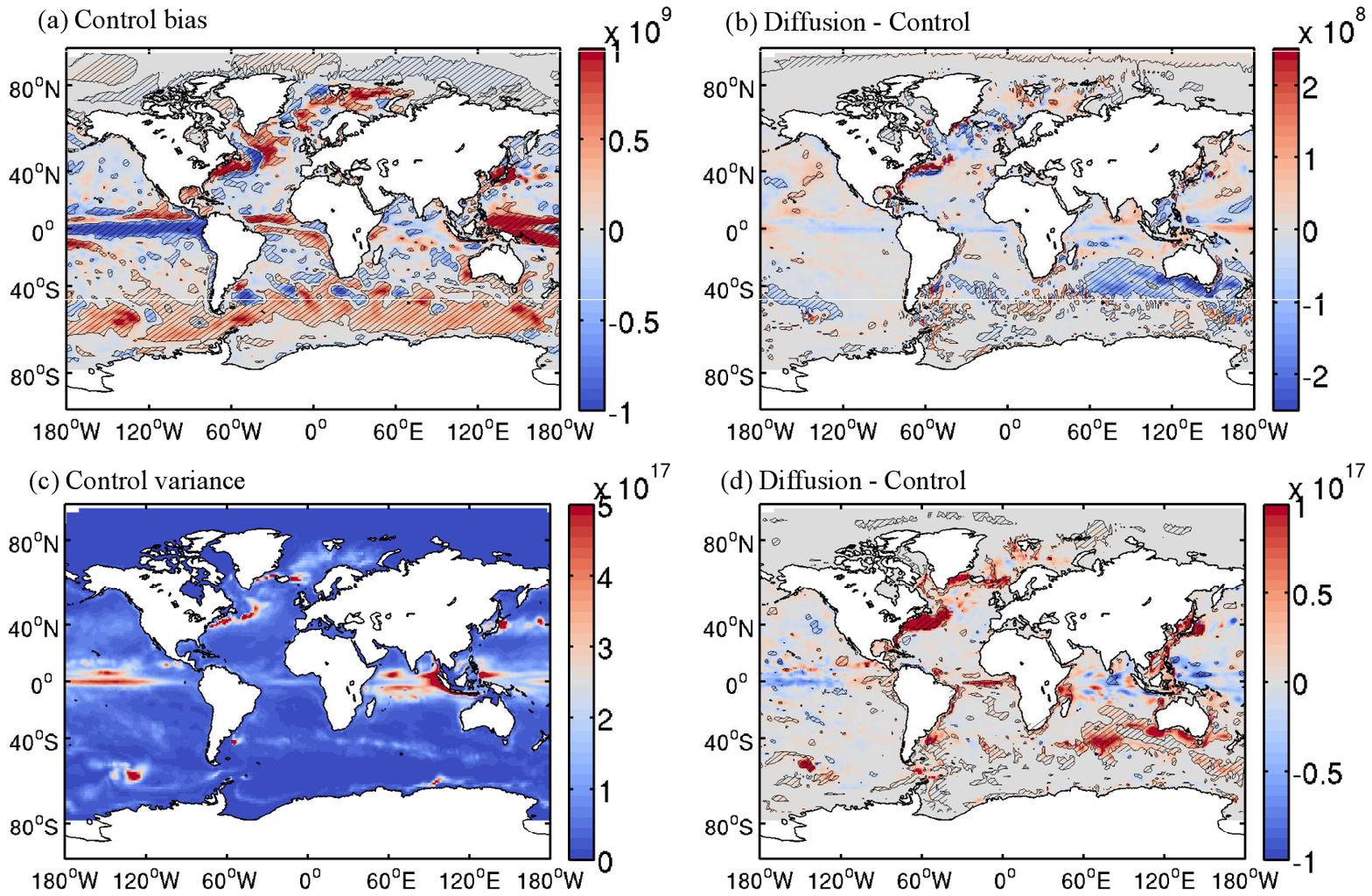}
\caption{Statistics of the upper 300m ocean heat content {\rm (J
    m$^{-2}$)} taken over days 60-90 of the seasonal forecast
  integration. (a) Mean bias in the control integration. (b)
  Difference in the bias between the integration with Stochastically
  Perturbed Parametrization Tendency (SPPT) and the control. (c)
  Ensemble spread (variance) in the control integration. (d)
  Difference in the ensemble spread between the integration with SPPT
  and the control. Hatched areas indicate regions that are
  distinguishable from zero using equation
  (\ref{significance:eqn}). Note the different colour scales.}
\label{maps:fig}
\end{figure*}
Seasonal forecasting with coupled atmosphere-ocean-land surface models has become well established at many numerical weather prediction (NWP) and climate forecast centres during the last two decades \citep[][]{MacLachlan14, Molteni11, Saha14}. These coupled systems for seasonal forecasting exploit predictability originating from the ocean and the land surface. More specifically, coupled models allow for predictions of seasonal to interannual variations of the climate system such as the ENSO (El Ni\~no Southern Oscillation) cycle, which may in turn affect predictions of other long time scale variations. Coupled models are also increasingly being used for medium-range weather predictions (e.g. run operationally at ECMWF since Nov 2013).

Forecast uncertainties need to be accounted for in these prediction systems. For example, errors in the observations and an incomplete observing system lead to inaccuracies in the initial conditions of forecasts. Due to the chaotic nature of the atmosphere and the Earth system, these initial errors tend to grow quickly, reducing the predictive skill of the forecast models. Ensembles of model simulations are used to account for initial condition uncertainty. Each ensemble member is initialized with slightly different initial conditions, generated using a perturbation method (e.g. ensembles of data assimilations, singular vectors). The ensemble as a whole then provides a probabilistic forecast. 

In addition to the uncertainties in initial conditions, forecast models themselves are inaccurate due to the numerical approximations used in the temporal and spatial discretization or due to inaccurate representations of sub-grid scale processes. In the past few decades, different strategies have emerged to account for these model uncertainties. Multi-model ensembles make use of the diverse set of weather and climate models to sample the uncertainty in the model formulation, on timescales of weather \citep[e.g.,][]{Mylne02, Park08}, seasonal \citep[][]{Palmer04, Weisheimer09} and climate \citep[e.g.,][]{Tebaldi07, Flato13} predictions. Perturbed parameter ensembles, on the other hand, sample the uncertainty in the choice of specific, imperfectly constrained model parameters. Furthermore, in multi-parametrization ensembles the set of applied parametrization schemes is modified for each ensemble member.
Finally, stochastic parametrizations have become a well-established technique for representing model uncertainty, especially for atmospheric NWP \citep[e.g.,][and references therein]{Palmer09}. By injecting stochastic perturbations into the system at the sub-grid scale, uncertainties in closure schemes are incorporated and unresolved sub-grid scale variability may be taken into account. Some of the stochastic schemes such as the Stochastically Perturbed Parametrization Tendency (SPPT) scheme have been used successfully in atmospheric forecast models \citep[e.g.,][]{ Palmer09}, leading to improved forecast skill up to the seasonal time scales \citep[][]{Weisheimer14}. The study of \citet{Weisheimer11} compared the multi-model, perturbed parameter and atmospheric stochastic physics approach for monthly and seasonal forecasts showing the potential for stochastic parametrizations to outperform the multi-model ensemble.

The motivation for stochastic parametrizations in the atmosphere originates to some degree from the existence of power law structures and the related rapid upscale error propagation \citep[see][for a detailed discussion]{Palmer12}. Since similar power law structures, associated with mesoscale eddies, can also be found in the ocean \citep[][]{LaCasce03, LaCasce08} similar arguments for the potential role of stochastic parametrizations may therefore hold.

Stochastic parametrizations have not only been introduced into the atmospheric component of NWP and seasonal forecast models \citep[e.g.,][]{Buizza99, Palmer09}, but have more recently also been implemented into the sea ice \citep[][]{Juricke13, Juricke14}, ocean \citep[e.g.,][]{Brankart13, Brankart15}, land surface \citep[][]{MacLeod15} and air-sea coupling  \citep[e.g.,][]{Williams12, BalanSarojini09} components of global general circulation models, to account for the uncertainty in sub-grid parametrizations. Although the impact of ocean stochastic parametrizations has been demonstrated in a climatological context \citep[e.g.,][]{Brankart13}, as yet no comparable stochastic parametrization in the ocean has been implemented into operational coupled models. Since some of the proposed stochastic schemes for ocean sub-grid scale parametrizations \citep[e.g.,][]{Mana14} may not be straightforward to implement in complex coupled global models, the purpose of this study is to investigate the impact of simpler stochastic schemes of similar complexity to those used in the atmosphere.

Three methods of stochastic parametrization are considered: A surface
flux parametrization similar to \citet{Williams12}, a stochastic
perturbation of the equation of state similar to \citet{Brankart13},
and stochastic perturbations of the parametrized tendencies of
diffusion, mixing and viscosity, which can be considered the
application to the ocean of the SPPT scheme used successfully in atmospheric
models \citep[e.g.,][]{Palmer09}. The details of the model setup, the
stochastic parametrization schemes and the methods of data analysis
are given in section \ref{model:sect}. The results of several ensemble
integrations over a 10 year period are presented in section
\ref{results:sect} and our conclusions are summarised in section
\ref{conclusions:sect}.

\section{Model Configuration and Experimental Design}
\label{model:sect}

\begin{figure*}
\noindent\includegraphics[width=40pc]{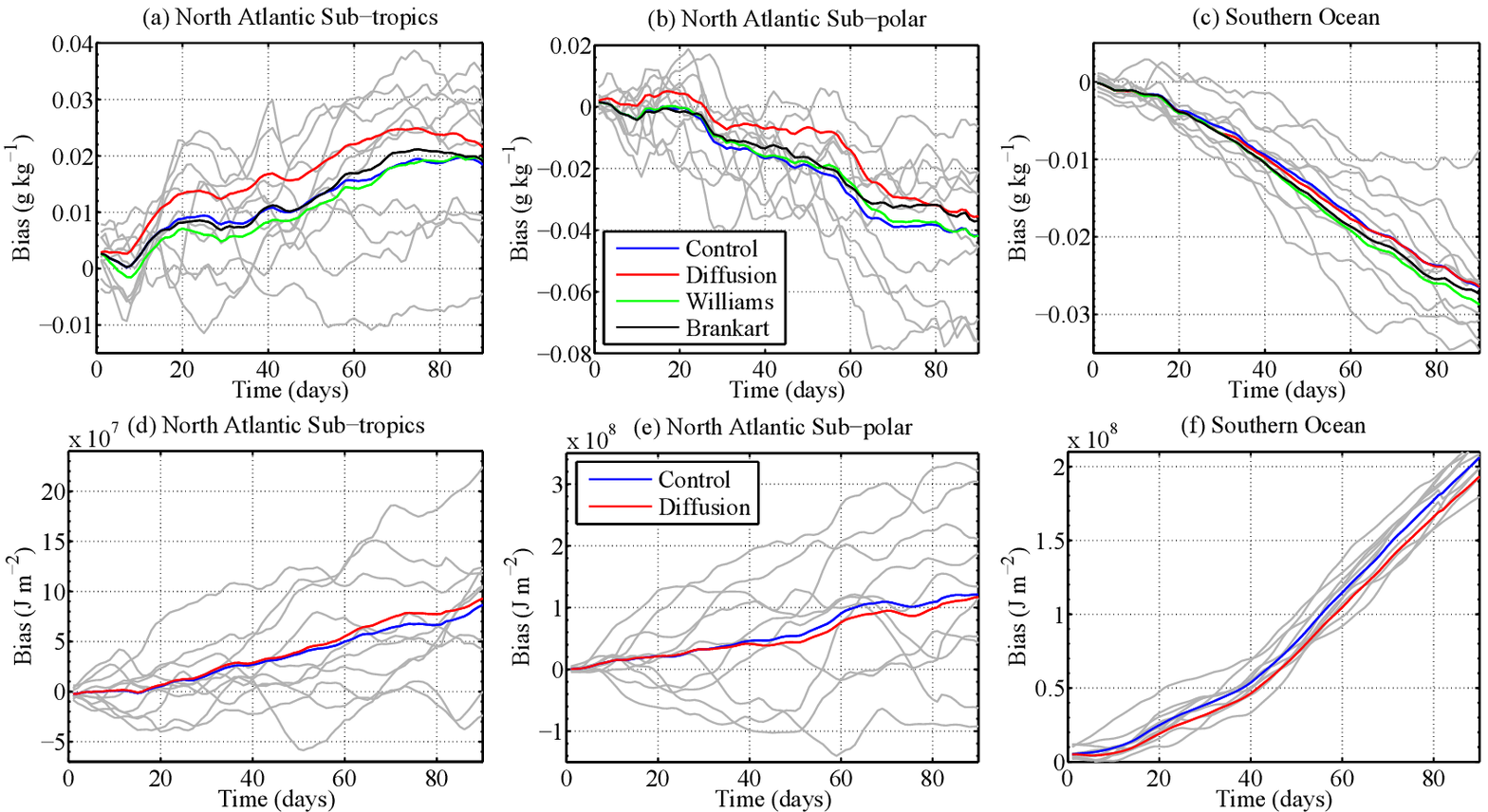}
\caption{Bias in the sea surface salinity (a), (b), and (c), and upper
  300m ocean heat content, (d), (e), and (f)), averaged over three
  regions (see text for details). The coloured lines indicate the
  ensemble mean bias averaged over all 10 start years for the control
  integrations (blue), the integrations with SPPT (red), SSF (green)
  and SES (black). The grey lines represent the ensemble mean bias of
  the control integration for each of the 10 years.  Note the
  different $y$ axis scales.}
\label{bias:fig}
\end{figure*}

Integrations were performed using a variation of the ECMWF seasonal
forecast system\footnote{Note to referees: This model was integrated
  on the IBM super computer at ECMWF which has now been switched
  off. This version of the model is therefore no longer available for
  use, however a new version of the seasonal forecast system is
  expected to be available on the new CRAY super computer at ECMWF at
  some point in 2015. (Comment to be removed from the final paper.)}
(System 4, \citet{Molteni11}). The ocean component consists of the
NEMO v3.0 global ocean primitive equation model \citep{Madec08}
discretised onto an approximately one degree ORCA tri-polar grid
\citep{Madec96} with 42 vertical levels. NEMO is coupled to the ECMWF
atmospheric forecast model IFS Cy36r4, integrated at T159 horizontal
resolution (reduced from T255 in System 4) with 91 vertical
levels. Ocean reanalysis data, used for initial conditions and
verification, was provided by the ECMWF operational ocean reanalysis
system ORAS4 \citep{Balmaseda13}.  The ensemble of 5 re-analyses is
driven by sampling uncertainty in winds and in deep ocean initial
conditions, and sub-sampling observation coverage. The ocean analyzes
are then augmented by applying SST perturbations with an associated
sub-surface temperature signal \citep{Molteni11}. The unperturbed
ensemble member of ORAS4 is used for verification. Atmospheric initial
conditions are derived from the ERA-interim data set \citep{Dee11},
without initial nor stochastic perturbations applied. To introduce
spatial correlation in two of the three oceanic stochastic schemes
tested, the horizontal domain was divided into a regular $30^{\rm{o}}
\times 30^{\rm{o}}$ latitude longitude grid. In each of these grid
cells a single uniformly distributed random number was generated at a
given time interval (1 or 30 days) and applied uniformly to each of
the ocean model grid cells within the $30^{\rm{o}} \times 30^{\rm{o}}$
region.

A 10-member ensemble of experiments, initialised on the 1st of
November for each of the 10 years 1989-1998, was integrated for 3
months. We define three regions for the purpose of summarising the time dependent nature
of the integrations: 1) The North Atlantic subpolar region, $45^{\rm
  o}$N $\rightarrow 65^{\rm o}$N, $70^{\rm o}$W $\rightarrow 0^{\rm
  o}$E. 2) The North Atlantic subtropics region, $10^{\rm o}$N
$\rightarrow 45^{\rm o}$N, $70^{\rm o}$W $\rightarrow 0^{\rm o}$E. 3)
The Southern Ocean, $35^{\rm o}$S $\rightarrow 65^{\rm o}$S, all
longitudes. These regions were chosen in order to focus on the areas
where the largest biases in the mean and variance occur. These are
also regions where the schemes examined here have the largest
impact. Results of area averages over other regions support the
findings described here, or are inconclusive (i.e. cannot be
distinguished from the noise). Here, our criteria is defined as
\begin{equation}
 \frac{m }{\sqrt{4\sigma^2/n} } > 1,
 \label{significance:eqn}
\end{equation}
where $m$ is the ensemble mean difference between a stochastically perturbed and a
control integration, $\sigma$ is the ensemble standard deviation of
the difference and $n=10$ is the number of independent samples,
assuming that the model state is independent from one year to the
next. Equation (\ref{significance:eqn}) represents the mean as a
fraction of the uncertainty in the mean and is similar to the 95\%
confidence two tailed t-test. The choice of a $2 \sigma$ threshold is
somewhat arbitrary and provides a rough guide. See Figure
\ref{statistics:fig} for an impression of the spread indicated by
(\ref{significance:eqn}).

\subsection{Stochastic Surface Flux (SSF)}
\label{Williams:sect}

Stochastic perturbations of the air-sea fluxes were applied to the
seasonal forecast model based upon the method described in
\citet{Williams12} who performed experiments with a lower resolution
coupled atmosphere-ocean GCM. Their ocean model (OPA 8.2
\citet{Madec98}) was integrated using the ORCA2 grid, which has
approximately a 2 degree horizontal resolution, with 31 vertical
levels. Their atmospheric model (ECHAM 4.6, \citet{Roeckner96}) was
integrated using a T30 spectral grid with 19 vertical levels. In
\citet{Williams12}, the air-sea fresh water flux $\Delta S$ and
non-solar heat flux $\Delta T$, were perturbed separately in two
separate experiments, given the following perturbation scheme:
\begin{equation}
\Delta T \rightarrow \left( 1+r_T \right) \Delta T,  \quad \text{  and  } \quad
\Delta S \rightarrow \left( 1+r_S \right) \Delta S
\label{WilliamsRandomPert:eqn}
\end{equation}
with $r_T$ and $r_S$ being a random numbers uniformly distributed between $\pm0.5$
and generated at 3 hours intervals on the ORCA2 grid scale. For this
paper, in addition to moving to a higher resolution, the interval over
which random numbers were chosen was increased to 1 day and applied
using the $30^{\rm{o}} \times 30^{\rm{o}}$ grid defined above. Also,
both fluxes were perturbed simultaneously but with different sequences
of random numbers.

\subsection{Stochastic Equation of State (SES)}
\label{Brankart:sect}

The method of stochastic parametrization of the nonlinear equation of
state, which relates the density to the temperature, salinity and
pressure, is based upon the method described in \citet{Brankart13}.
To simulate the uncertainty related to area-averaged temperature and
salinity fields used as input for the equation of state, a first order
auto-regressive process perturbs both state variables by an amount
proportional to their gradients. The auto-regressive process has a
decay timescale of 12 days, changed to 7.5 days in our
integrations. Ultimately the density at each grid point is perturbed
independently of neighbouring grid cells. In \citet{Brankart13}
integrations were performed using NEMO with the same ORCA2 grid as
that used by \citet{Williams12}. Their system, however, was forced
using climatological atmospheric data without inter-annual variations
\citep{Large09} rather than the atmosphere from a coupled model.

\subsection{Stochastically Perturbed Parametrization Tendency (SPPT)}
\label{stochasticDiffusion:sect}

We introduce a Stochastically Perturbed Parametrization Tendency
(SPPT) scheme for the ocean similar to that implemented in atmospheric
models for ensemble weather forecasts. The subgrid-scale parametrized tendencies (used to crudely mimic turbulent diffusion, mixing, convection, viscosity) applied to the zonal
velocity, $u$, meridional velocity, $v$, salinity, $S$, and
temperature, $T$, are multiplied by $(1+r_X)$ as in
(\ref{WilliamsRandomPert:eqn}), with different random sequences $r_X$
for $u$, $v$, $S$, and $T$, i.e. $X\in\{u,v,S,T\}$.  For example, the deterministic prognostic equation for $T$ given by
\begin{equation}
  \frac{\partial T}{\partial t}= -\nabla \cdot \left(T \mathbf{U}+T \mathbf{U}_{\text{GM}}\right)+D_T+F_T
 \label{temperature:eqn} 
\end{equation} 
takes the following form when SPPT is implemented
\begin{equation}
  \frac{\partial T}{\partial t}= -\nabla \cdot \left(T \mathbf{U}+T \mathbf{U}_{\text{GM}}\right)+(1+r_T) D_T+F_T,
\end{equation} 
where $\mathbf{U}=(u,v,w)$ is the 3D Eulerian velocity,
$\mathbf{U}_{\text{GM}}$ is the eddy-induced velocity from the
Gent-McWilliams parametrization scheme \citep{Gent90}, $D_T$
represents the parametrized diffusion and mixing tendencies and $F_T$
the air-sea flux.

Summarised in the respective $D_X$ terms for the momentum and tracer
equations are parametrized terms using vertical eddy viscosity and
diffusivity coefficients calculated by a turbulent kinetic energy
(TKE) closure scheme as well as a double diffusion mixing scheme.
Lateral diffusion and viscosity are also included in $D_X$ using
horizontally varying coefficients for tracers (following
\citet{Held96}, and \cite{Treguier97}) as well as a three dimensional
spatially varying viscosity coefficient for the momentum equations.
See \citet{Madec08} for the specification of these schemes.  The
spatial field of the random numbers $r_X$ was generated at either 30
day or alternatively 1 day intervals using the $30^{\rm{o}} \times
30^{\rm{o}}$ grid defined above with the same values of $r_X$ applied
on each model level. The four $r_X$ were applied to the parametrized
tendencies of the respective prognostic equations simultaneously for
all four fields and were drawn from a uniform distribution
between $\pm 0.8$. This value for the magnitude of $r_X$ was found to
be the maximum value consistent with model stability.

\section{Impact of the Stochastic Parametrizations on Model Bias and
  Ensemble Forecast Performance}
\label{results:sect}

The control integrations without any stochastic perturbations exhibit
bias relative to the reanalysis. The daily bias is estimated by taking
the difference between the reanalysis and the mean of each ensemble
mean over all ten start years. Averaging in time between 60 and 90
days into the integration indicates that this bias, as illustrated for
upper 300m ocean heat content in Figure \ref{maps:fig}(a), often
coincides with the regions of high variability such as the eastern
tropical Pacific, Southern Ocean, Gulf stream and Kuroshio regions.
The upper 300m heat content is chosen as it yields a particularly
strong signal to noise ratio compared to sea surface temperatures in
isolation.

The SPPT scheme leads to a change in the bias in the Southern Ocean,
particularly in the region of the south coast of Australia, and the
North Atlantic (Figure \ref{maps:fig}(b)). Comparing panels (a) and
(b) in Figure \ref{maps:fig} indicates that the warm bias along the south
coast of Australia is reduced. Taking the area means, specified in
section \ref{model:sect}, of the daily bias (Figure \ref{bias:fig})
highlights a reduction in the mean bias for the upper 300m ocean heat
content and sea surface salinity (SSS) in the North Atlantic subpolar
region after the first month. The bias is initially not exactly zero
due to the random spread in the initial condition perturbations. In
the Southern Ocean, while the warm bias has been reduced, no
noticeable changes are observed in the SSS bias. In contrast, the bias
in the North Atlantic subtropical region has increased due to SPPT.
For certains regions the difference in bias remains relatively constant over the length of the integration.

The grey lines in Figure \ref{bias:fig} represent the ensemble mean
bias of the control integration for each of the 10 years and can be
considered as an indicator of confidence. Changes in the ensemble bias
and spread (defined as the ensemble variance) between the control and
integrations using the SSF and SES schemes are too small to
distinguish from zero.  The inherent ensemble spread due to the ocean
initial condition perturbations and atmospheric variability is large
in comparison.  When it comes to the area-averaged quantities, Figure
\ref{bias:fig} also confirms that the SSF and SES schemes have little
impact upon the bias though there are regions in which the bias
appears slightly reduced (e.g., North Atlantic subpolar SSS around 2-3 months for the SES scheme).

\begin{figure*}
  \noindent\includegraphics[width=40pc]{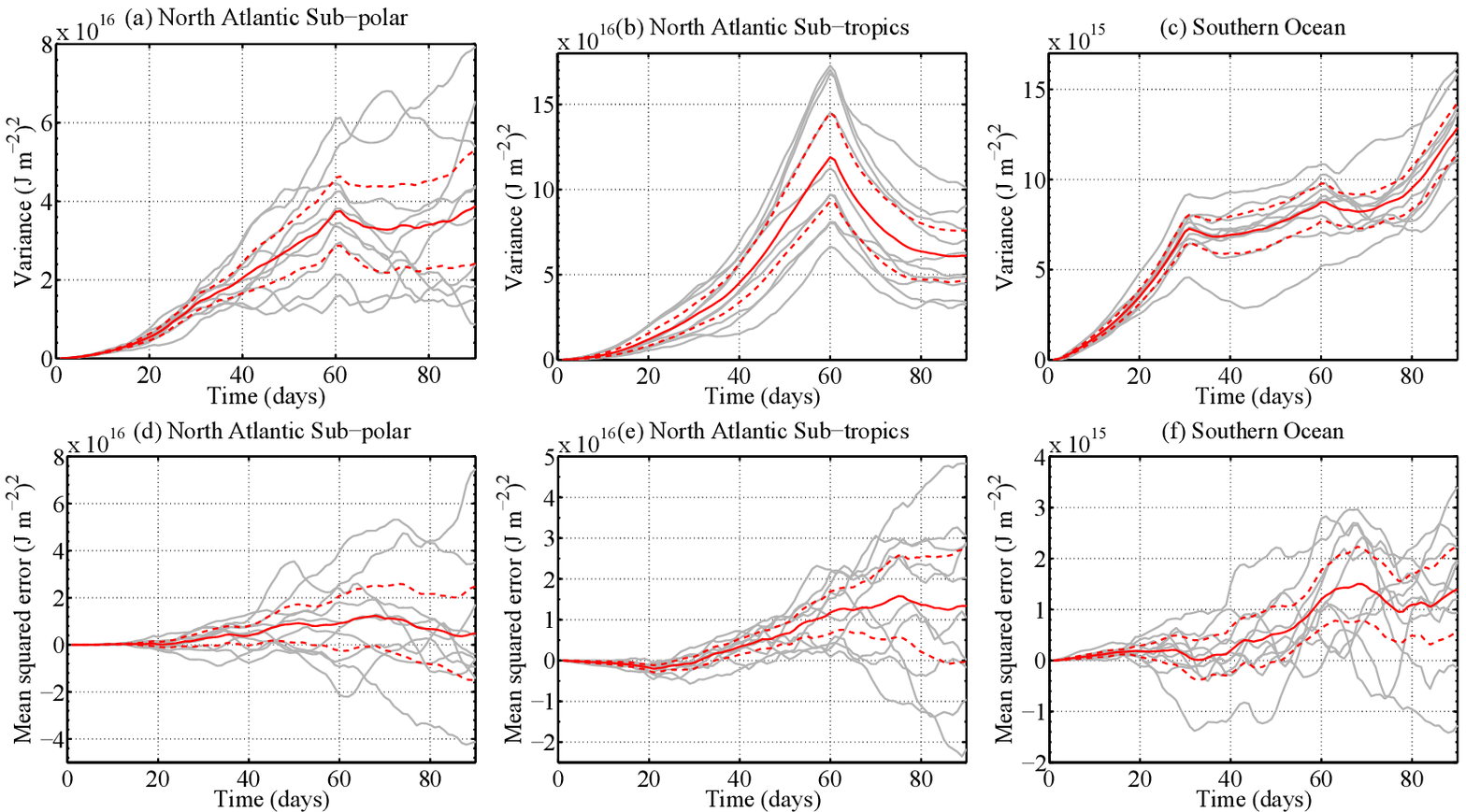}
  \caption{Area averaged changes in the ensemble spread (top row) and
    mean square error (bottom row) with respect to the control
    integration, due to SPPT, of upper 300m ocean heat content, see
    text for details. The grey lines indicate the statistics from the
    ensemble taken over each of the 10 start years. The solid red line
    indicates the mean over all years and the dashed lines indicate
    the uncertainty in the mean (the mean plus or minus two standard
    deviations divided by $\sqrt{10}$).}
\label{statistics:fig}
\end{figure*}

The impact of the SPPT scheme upon the ensemble spread is apparent in
several key regions (Figure \ref{maps:fig}(d)). The largest and most
significant impact is shown in the region of the south coast of
Australia with patches of visibly increased spread throughout the
Southern Ocean. This is a similar pattern to the changes in vertical
mixing induced by a stochastic wind forcing compared to climatological
wind forcing \citep[][compare with their Figure
8]{BalanSarojini09}. In addition to the Southern Ocean, Figure
\ref{maps:fig}(d) also reveals increased ensemble spread along western
boundary currents, including the Gulf stream and Kuroshio.  Although
there are reductions of ensemble spread in parts of the tropics, they
are not distinguishable from zero using the criteria
(\ref{significance:eqn}). The increase in ensemble spread over the
length of the integrations is readily seen in area means of the North
Atlantic subtropical and subpolar regions and the Southern Ocean
(Figure \ref{statistics:fig} a-c). The sharp changes in spread
apparent in Figure \ref{statistics:fig} (a), (b) and (c), is due to a
new set of random numbers being chosen every 30 days. We expect that
substituting the $r_X$ for auto-regressive processes with the equivalent
timescales will remove this artefact.

On the other hand, Figure \ref{statistics:fig} d-f indicates that the
SPPT scheme has increased the regional spread at the expense of
increased forecast error in the upper ocean heat content in most
regions over the length of the integration.  The forecast error is
defined as the difference between the bias corrected ensemble mean and
the reanalysis.  Bias corrected means that the daily bias is
subtracted from each integration. The mean squared error is then the
squared error, averaged over the region of interest. The forecast
error is slightly reduced between 2 and 5 weeks in the North Atlantic
subtropical region. Comparing figures \ref{statistics:fig} (a) with
(d), (b) with (e) and (c) with (f), indicate that the additional mean
squared error is approximately an order of magnitude lower than the
increase in spread.  The equivalent results for the SSF and SES
schemes show that there is little impact upon the total ensemble
spread.

Reducing the time between random numbers from 30 days to 1 day
resulted in a reduced impact for the SPPT scheme (to the point at
which changes are almost indistinguishable from zero) when comparing
ensemble bias, spread and mean squared error to those of the control
integration.  This supports the conventional hypothesis that the
impact of the stochastic term is strongly dependent upon its
timescale, with longer timescales corresponding to a larger impact.
Integrations in which the spatial correlation of the noise is reduced
to a $20^{\rm{o}} \times 20^{\rm{o}}$ degree grid yield only small
changes. We would expect that as the correlation length scale is
reduced further, there will come a point at which the impact of the
random term is strongly reduced (see for example
\cite{Juricke13}). Additional
figures were produced (not shown here) demonstrating the reduced impact at these timescales and
equivalent plots for the SSF and SES schemes and the impact on the sea
surface temperature and salinity.

\section{Conclusions}
\label{conclusions:sect}

In this study, we test three oceanic stochastic parameterization
schemes: Stochastically Perturbed Parametrization Tendency (SPPT),
Stochastic Surface Flux (SSF) \citep{Williams12} and a Stochastic
Equation of State (SES) \citep{Brankart13}. The relatively simple SPPT
scheme, which has proved an important element of an ensemble forecast
system in numerical weather prediction, injects multiplicative noise
into the prognostic equations with an amplitude proportional to the
deterministically parametrized tendencies. These three schemes are
applied to the ocean component of a state-of-the-art seasonal coupled
forecast system and account in part for the uncertainty in sub-grid
processes.  The model considered here exhibits relatively large
oceanic variability compared to a system run at coarser resolution or
without an interactive atmosphere. Using such a model, the impact of
the SSF and SES schemes was relatively small at the monthly timescales
considered, and will likely be more visible through longer timescale
integrations.  In the case of the SES and SPPT schemes, the amplitude
of the stochastic perturbations was limited by model stability.

Results show that compared with the other schemes, SPPT is an
effective stochastic parametrization for increasing ensemble spread
for variables such as sea-surface temperature and salinity, and upper
300m ocean heat content. The impact of SPPT was found to be
particularly marked, and visible above the background variability, in
regions of strong eddy activity, such as along western boundary
currents in the Gulf Stream and Kuroshio regions, in the North
Atlantic sub-polar region and also in parts of the Southern Ocean.

On the other hand, ensemble-mean forecast skill was not improved by
the addition of SPPT, except for in the North Atlantic sub-tropics,
and for some regions model bias was made worse. The latter does not
necessarily imply that SPPT is unrealistic, since the value of many
climate model parameters are found by running the model in
deterministic mode and estimating the values that fit the observations
best. If a stochastic scheme impacts on the model mean state, then
such tuning should be performed using the full stochastic model and
not a deterministic approximation to it \citep{Palmer12}. In addition
to tuning the deterministic parameters of the model, the impact of a
stochastic parametrization may be tuned by adjusting the decorrelation
timescale and spatial distribution of the random perturbations as well
as their magnitude. For our particular configurations, changes to the
timescale appeared more important than the spatial scale.

The fact that we have not been able to reduce ensemble-mean forecast
error, even when the model fields have been bias corrected
a-posteriori, suggests that SPPT may be too crude a scheme for ocean
models. In the ocean, sub-grid processes are parametrized largely by
diffusion and viscosity. The impact of uncertainty in these terms upon
the large-scale oceanic circulation, as simulated by SPPT, may not
well reflect the uncertainty in the underlying turbulent
processes. What this suggests is that a more positive impact than has
been found here requires the development of more sophisticated
stochastic schemes for unresolved and missing processes.

The development of such stochastic closures should when possible
remain consistent with fundamental physical constraints. For example,
our initial implementations of a stochastic Gent-McWilliams scheme
violated important non-divergent and adiabatic constraints leading to
unrealistic upwelling and growing instabilities occurring as soon as
the stochastic forcing is increased beyond a certain amplitude. Avenues for future investigations might
include a more sophisticated Gent-McWilliams scheme that has a
large impact, is stable over long time periods, and remains
consistent with the non-divergent and adiabatic constraints of the
deterministic scheme.  New stochastic schemes should ultimately be
guided by observations, however ocean observations are sparse.  As a
methodology to develop new parametrizations, high-resolution idealised
simulations can be substituted as "truth"
\citep[e.g.,][]{Berloff05,Berloff15} and optimally coarse-grained to
derive stochastic terms which can for example be decoupled from the
background flow \citep[e.g.][]{Cooper15b}.  A complementary approach
is to implement a PDF-based parametrization, such as \citet{Mana14}
who have developed a stochastic parametrization of ocean mesoscale
eddies which depends on the temporal tendency of potential vorticity.

\begin{acknowledgments}
  This work was funded by UK NERC grant NE/J00586X/1, along with a
  generous allocation of super-computer resources from ECMWF. AW was partly supported by the EU project SPECS funded by the European Commission Seventh Framework Research Programme under the grant agreement 308378. Thanks
  go to Peter D\"{u}ben and Aneesh Subramanian for useful discussions and
  encouragement during the preparation of this paper and to J.M. Brankart for sharing the code associated with his parametrization. The model data can be obtained by emailing Stephan Juricke, Stephan.Juricke@physics.ox.ac.uk.
\end{acknowledgments}

\bibliographystyle{agufull08} 
\bibliography{database,references}

\end{article}

\end{document}